**River interlinking alters land-atmosphere feedback and changes the Indian summer monsoon.**

Tejasvi Chauhan[1], Anjana Devanand[2,3,4], M. K. Roxy[5], Karumuri Ashok[6], Subimal Ghosh[1,2,*]

1. Department of Civil Engineering, Indian Institute of Technology Bombay, Mumbai, India
2. Interdisciplinary Program in Climate Studies, Indian Institute of Technology Bombay, Mumbai, India
3. Australian Research Council Centre of Excellence for Climate Extremes, University of New South Wales, Sydney, NSW, Australia
4. Climate Change Research Centre, University of New South Wales, Sydney, NSW, Australia
5. Centre for Climate Change Research, Indian Institute of Tropical Meteorology, Pune, India
6. Centre for Earth, Ocean and Atmospheric Sciences, University of Hyderabad, Hyderabad, India

[*]Corresponding Author. Email: subimal@iitb.ac.in

## Abstract

Massive river interlinking projects[1–4] are proposed to offset observed increasing trends of extremes[5–10], such as droughts[11,12] and floods[13] in India, the second highest populated country. These river interlinking projects[3] involve water transfer from surplus to deficit river basins through reservoirs and canals, but without an in-depth understanding of the hydro-meteorological consequences. Using information theory-based causal delineation techniques[14–17], a coupled regional climate model[18,19], and multiple reanalysis datasets[20,21], we show that causal pathways exist across different basins in India due to strong land-atmosphere feedback, which disputes the generally practiced assumption of hydrological independence between river basins. The causal information from one basin's land crosses the basin boundary through the atmosphere. We further find that increased irrigation from the transferred water reduces mean rainfall in September by up to 12% in many parts of India most of which are already water stressed. We observe more drying in La Niña years as compared to El Niño years. Reduced September precipitation can lead to drying of rivers post monsoon augmenting the water stress across country rendering interlinking dysfunctional. These findings demand model-guided impact assessment studies of large scale hydrological projects across the globe considering land-atmosphere interactions.

**Main**

Large international rivers, such as the Ganga, the Brahmaputra, and the Indus, are central to the development of agriculture-dominated India, with a 1.4 billion population[22]. However, like river basins around the globe[23], Indian river basins are also under severe stress due to global climate change[24], massive population growth[24–26], increased uncontrolled human water use[27,28], and pollution[29,30]. Indian summer monsoon rainfall (ISMR) during June to September is the primary source of water in Indian river basins and accounts for almost 80% of the country's annual rainfall and governs the Gross Domestic Product (GDP)[31]. Over the last few decades, ISMR has experienced a decline in the mean rainfall[32,33] and an increase in the intensity, occurrences, and spatial variability of extreme rainfall[5,6,8,34]. Such changing meteorological patterns have resulted in increased hydrologic extremes, floods, and droughts in India[6,10–12,35–37]. These hydrologic changes have augmented the water stress across the country, elevating the risk of disasters. As an adaptation measure to combat the increasing hydrologic extremes, India has planned massive interlinking projects on its rivers with a proposed budget of USD 168 billion[1,2,4]. The proposal[2,3] involves a network of canals with an approximate length of 15,000 km and 3000 reservoirs with the capacity to transfer 174 billion cubic meters of water each year from surplus to deficit basins and generate 34 million kilowatts of hydropower along with benefits[38] like flood control, drought mitigation and navigation. The experiences of river interlinking in China showed the stabilization of groundwater[39]; however, such an ambitious plan may significantly impact the ecology[40] of the aquatic ecosystem and fish diversity[41]. Literature also showss that increased river regulation could increase the water foot print[28]. Hence, interlinking must be carefully designed so as to optimize the conflicting objectives of ecological sustainability and meeting water demands[42].

Critically, the interlinking of the rivers assumes the river basins to be hydrologically independent because land water flow does not cross the basin boundary. So far, no scientific studies have explored the possibility of feedback of the inter-basin water transfer to the water cycle. We hypothesize that the water transfer may impact the donor or adjacent basins through land-atmosphere feedback. Such possibilities could be high in the Indian region, where the land feedback to the atmosphere is also high[43,44]. In the present study, we test the hypothesis by developing a causal network between the atmosphere and land variables across river basins in India. We use information theory-based transfer entropy (TE)[17,45] and a causal network learning algorithm called Peter Clark's Momentary Conditional Independence (PCMCI)[14] (See methods for details) to generate causal networks between different hydrometeorological variables. The TE is an information-theoretic metric that finds the dependence between two variables by excluding the effects from the history of the target variable. It can capture the nonlinear and lagged causal connections between variables, and can be considered as a nonlinear extension of Granger Causality[46]. The TE becomes difficult to estimate for a large number of variables because of the complexity due to dimensionality. The PCMCI adopts a two-step method to handle the curse of dimensionality while delineating causal structure from time series (methods).

We use both the approaches (TE and PCMCI) on the variables (Extended Table 1), soil moisture (SM), latent heat flux (LH), sensible heat flux (SH), precipitation (P), relative humidity (R), wind speed (WS) (resultant of u-wind (U), v-wind (V)), incoming shortwave radiation (SR), and temperature (T) over the major river basins of India (Figure 1(a)), Ganga (G, 808334 km$^2$), Godavari (Go, 302063 km$^2$), Mahanadi (M, 139659 km$^2$), Krishna (K, 254743 km$^2$), Narmada-Tapi (NT, 98,796 km$^2$, 65,145 km$^2$ respectively – two river basins taken together), and Cauvery(C, 85624 km$^2$). Since Narmada and Tapi basins are relatively small, we club them together and represent them as a single basin. Variables relative humidity, u-wind, and v-wind are taken at 850 hpa pressure level and remaining variables are near surface. We first use 40 years (1980-2019) of daily reanalysis data from the European Center for Medium-Range Weather Forecast (ECMWF, ERA-5)[20]. The extended Figure 1 presents the climatology of soil moisture, precipitation, and Evapotranspiration (ET, generated from latent heat flux) for different basins generated using ERA-5 variables. All the basins receive maximum precipitation during the Indian summer monsoon (also called southwest monsoon) from June to September[47]. The Cauvery basin also receives significant amount of rainfall during October-December during the northeast monsoon[48] and has two peaks in annual precipitation. The soil moisture in Ganga, Godavari, Krishna Mahanadi, and Narmada-Tapi basin peaks during August and starts declining by the end of the summer monsoon season. The soil moisture in Cauvery peaks during late October, showing cumulative effects of rainfall from Indian summer monsoon and northeast monsoon. Evapotranspiration in all basins increases during the start of the Indian summer monsoon and is highest during post-monsoon because of moisture accumulation during monsoon getting exposed to solar radiation.

After performing causal analysis, we represent the association between variables across different basins as networks. We demonstrate the causal relationships between land variables across basins through land-atmosphere, atmosphere-atmosphere, and atmosphere-land interactions showing that the basins are not hydrologically independent. A perturbation in a river basin due to the proposed interlinking can travel to the neighboring basins by atmospheric pathways. Further, we used a modified regional climate model — Weather Research and Forecast coupled with Community Land Model 4 (WRF-CLM4, details in Methods)[18] — to test the hypothesis that by land-atmosphere feedback, the additional irrigation from river interlinking can lead to changes in the Indian summer monsoon spatial patterns and the hydrology of the neighboring basins. To the best of our knowledge, such feedbacks have not been considered in the literature for any globally existing or planned interlinking projects.

This study shows that river basins are linked to each other by land-atmosphere feedback and any perturbation in one basin can travel to neighbouring basins. This result is at odds with the general assumption of independence of river basins while planning hydrological projects. We also show that by land-atmosphere feedback, river interlinking projects in India will affect Indian summer monsoon leading to a reduction in September rainfall in dry regions of the country which can further aggravate the water stress. The methodology and results presented here pave way for similar scientific assessment of the impacts of interlinking for river basins and other large scale hydrological projects across the globe.

## Results and Discussion

### Information Links between the Basins

We computed the transfer entropy (TE, Details in Methods) between the land variables across the basins using 40 years of daily reanalysis data from ERA-5 (1981-2020). We used the entire period for deriving the TE with a maximum lag of 10 days. Figure 1(c) shows the directional bivariate TE links between land variables across basins showing a link only if it is found statistically significant at 95% confidence (see methods) . Each arc is a variable, and the arrow represents the link's direction to the other variable it projects onto. With the increase in the number of variables, estimating multivariate TE becomes computationally expensive; hence, we stick to bivariate analysis. We get a large number of links between land variables using TE across the basins. We found that the Ganga basin land variables produce a high number of causal links affecting land variables of other river basins with a low number of incoming links from other basins. For example, the latent heat flux and soil moisture of the Ganga basin shows the existence of causal links to at least one of the land variables of all other basins and there is no incoming link to soil moisture of Ganga basin from other basins. High recycled precipitation due to land-atmosphere feedback is well established for the Ganga basin[43,44,49]. Cauvery basin on the other hand has a large number of incoming links from all other basins, as evident from Figure 1 (c). Literature shows that the Cauvery basin receives recycled precipitation generated by evapotranspiration from the neighboring regions[50]. Stronger causal connections exist between the land variables of other basins as well.

The above bivariate TE analysis shows causal relationships across land variables; however, some of the links we see in the Figure 1 are likely due to common drivers such as the El Niño-Southern Oscillation (ENSO), and/or indirect links. A co-variability between the soil moisture in the Mahanadi basin with that of Ganga, for example, may be associated with monsoonal variations or the interannual impacts of ENSO. It is also possible that the local changes in soil moisture in a river basin may indirectly affect that in another river basin through local land-atmosphere interaction, which in turn would affect the large scale flow. Thus, the most likely reason for these links would be indirect links or common drivers. To address these challenges, we used an advanced causality approach, PCMCI (see Methods for details), which tries to account for common drivers while controlling the high-dimensionality. The monsoon drives the climate and the water cycle in India. To understand the land-atmosphere processes and its impacts on water cycle, we have performed analysis using PCMCI for the summer monsoon season. We have applied PCMCI to all the land and atmospheric variables from reanalysis data separately for each year's monsoon seasons with 122 days each, considering a maximum of 10 days' lag. Figure 2 presents the links appearing for more than 10 years out of 40 years (1981-2020). We hypothesized that the causal connections between land variables of two different basins A and B exist through a series of indirect links: land variable (river basin A) → atmospheric variable (river basin A) → atmospheric variable (river basin B) → land variable (river basin B). We present the links in the same way in Figure 2. The nodes in the leftmost column are the source land variables. The links from these nodes go to the second layers, the atmospheric variables of the same basin. For example, the link originating from the latent heat of Ganga (LH_G) goes to the temperature of Ganga (T_G). The 3rd vertical layer from the left presents the atmospheric variables of the information receiving basins. The land variables cannot cross the river basin boundary, but the atmospheric variables can. The links from the 2nd layer to the 3rd layer represent the inter-basin atmospheric links. For example, the Ganga basin temperature affects the temperature of the Mahanadi basin (T_M'). The "apostrophe" sign after a basin symbol (for example, M for Mahanadi in T_M') signifies the receiving basin's atmospheric variable. The 4th layer, or the rightmost layer, contains the land variable of the receiving basin. The temperature of the Mahanadi basin (T_M') impacts the soil moisture of the Mahanadi basin (SM_M). Hence the pathway between LH_G to SM_M is LH_G→T_G→T_M'→SM_M. Interestingly, we also see a similar pathway from the Mahanadi variable to that of Ganga, indicating a feedback.

The intra-basin land-to-atmosphere connection happens by evapotranspiration contributing to the moisture content of the air while causing surface cooling, whereas atmosphere-to-land connections occur by precipitation and temperature changing soil moisture and surface energy balance[45,51]. The atmosphere-to-atmosphere interactions between different basins occur through moisture and heat transported by winds across basin boundaries. To make sure that the links are not specific to a single reanalysis product, we applied PCMCI to another reanalysis data: Modern-Era Retrospective analysis for Research and Applications, Version 2 (MERRA-2)[21]. The results are shown in Extended Figure 2. We found similar links existing in the network derived from MERRA-2. There is a clear resemblance in the nodes present in the 4 layers across the reanalysis, though they are not exactly the same. This proves that, due to the land-atmosphere feedback, the assumption of hydrological independence of neighboring basins does not hold true.

**Feedback from the proposed interlinking**

Based on the above causal analysis, we hypothesize that a perturbation in the land variables of a receiver basin due to the proposed interlinking can also affect its neighboring basins (for example, feedback between Ganga and Mahanadi explained above), including the donor basin through feedbacks. To test the hypothesis with a physics-based model, we used the coupled land-atmosphere model WRF-CLM4 (Details in Methods). We have chosen the period of Indian Summer Monsoon (15 May to 31 October, initial 16 days from 15 May to 31 May used as spin up run every year) from 1991 to 2012 to see the potential impacts of surplus irrigation by interlinking projects on other basins (see Methods). Our control run (hereafter CTL) contains the currently practiced irrigation in India as obtained from the agricultural census data and has previously been demonstrated to posess a reasonable skill in simulating Indian summer monsoon[9]. We performed another simulation (the irrigation run, hereafter IRR) by increasing the percentage irrigated area to 80% in regions where interlinking projects target an increase in the culturable command area as shown in Figure 1 (b). Figure 1(b) shows the increase in the percentage irrigated area in each grid cell done to achieve 80% irrigated aria in IRR run. The IRR run provides irrigation (in addition to irrigation in CTL run) of around 600mm/day (1400mm/day for paddy[9]) to the area of around 30 million hectares across the country (Figure 1(b)). The simulations consider the India-specific crop and irrigation practices[9,52,53] (details in methods). The differences in results between the two simulations highlight the feedback from the interlinking through land-atmosphere interactions.

Extended Figure 3 shows difference in mean daily precipitation between IRR and CTL runs (IRR-CTL) for the monsoon season (b); June to September, JJAS), June (c), July (d), August (e), September (f). Hatched lines in plots indicate the regions where the difference is statistically significant at 90% confidence level. Spatially, there is not much statistically significant changes in precipitation during June, July, and August, though during July, we see some rainfall deficit in the western India and a slight increase in precipitation for the rest of the country. Importantly, September sees a widespread and maximum statistically significant reduction in precipitation. The simulated changes in September rainfall can be attributed to land-atmosphere feedback. Contribution of land-atmosphere feedback to Indian summer monsoon is known to peak during September when the soil contains high moisture leading to high evapotranspiration[30]. The mean daily precipitation during September shows a statistically significant reduction of up to 4 mm/day in central India and some parts of north and western India (Extended Figure 3, f). The mean monsoon rainfall shows minimal changes which are similar to September precipitation. Hence, we have selected September month for further analysis.

Figure 3 shows the percentage change in mean September rainfall for various regions (Figure 3 b-i) in India for all years, with the median value shown by the horizontal line in each plot. The regions which will experience declines in September monsoon rainfall are coastal Gujrat, dry western region of Rajasthan, western Himalayan foothills in Uttarakhand, central India, east peninsular India, and east-central India. The highest median reduction is around 12% in the east central India in the state of Odisha. The decrease in September rainfall due to interlinking reaches almost 20% in 7 out of 22 years for state of Odisha and Uttarakhand in eastern central India, and western Himalaya respectively. A decrease of upto 50% exists in 7 out of 22 years in the state of Gujarat in western central India. We observe very high reductions in the simulated September rainfall due to interlinking in the driest region of India, the dry western region of Rajasthan, with an almost 12% decline on average. The rainfall in central India, a part of the core monsoon zone, also shows a 10% decline in the simulated september rainfall due to interlinking. The western Himalayan foothills in Uttarakhand and east-central India also show a moderate decline in September rainfall due to excess irrigation from the proposed interlinking. It is worth noting that while there is a reduction in September precipitation in generally dry parts of country, there is also an increase in September precipitation upto 12% in east India (states of Bihar, Jharkhand, and eastern Uttar Pradesh) and some parts of the deccan plateau (states of Mahatashtra and Telangana).

What is more, the interlinking will also result in a changing spatial pattern of temperature over India (Extended Figure 4, a). Extended Figure 4 shows the difference in mean daily values for daily maximum temperature, surface latent heat flux(b), and root zone soil moisture (c) between IRR and CTL runs for September. The changing meteorological patterns result in statistically significant changes in mean monthly soil moisture and latent heat flux. However, there is a lack of one-to-one consistency everywhere due to complex hydrometeorological processes. The regions with less precipitation are accompanied by an increase in daily maximum temperatures of up to 1°C. and a decrease in soil moisture of around 15 mm (Extended Figure 4; a,c, respectively). The changes in soil moisture in the grids receiving daily irrigation cannot be used to quantify impacts of land-atmospheric feedback, and hence, masked with grey color. The irrigated grids are also visible to some extent as having high latent heat flux in Extended Figure 4(b). The proposed interlinking aims to improve the soil moisture everywhere by taking the surplus runoff from surplus regions and irrigating the deficit regions. However, contrary to this expectation, the feedback from the extra irrigation at the deficit basins results in declining rainfall in many neighboring areas, with a decline in soil moisture there. While there is a post-interlinking increase in soil moisture of Krishna basin, western portion of Godavari and Narmada-Tapi basins, and eastern Ganga basin, there is a pronounced decline in soil moisture of Mahanadi, Godavari, and western part of ganga basin in Indian states of Odhisa, Chhattisgarh, norther Maharsahtra, Madhya Pradesh, and Rajasthan. Hence, the purpose of improving soil moisture in deficit regions is met but at the cost of declining available water in the neighboring regions. Such unexpected feedbacks were unforeseen in the planning stage, and we contend that it is necessary to consider the land-atmosphere feedback processes in arriving at the policy decisions related to the interlinking.

El Niño Southern Oscillation (ENSO) is a major driver of interranual variability of the monsoon rainfall over India. To understand the interannual variations of the land-atmosphere feedback from the excess irrigation proposed through interlinking, we separately analysed El Niño and La-Niña years (Extended Table 2, Extended Figures 5 for El Niño and 6 for La Niña). These include both canonical and Modoki types of ENSOs, given that the impacts are qualitatively same[54]. We have produced the differences in mean daily values of precipitation (a), daily maximum temperature (b), root zone soil moisture (c), and surface latent heat flux (d) between IRR and CTL runs in both the extended figures for September. Interestingly, soil moisture drying due to excess irrigation is more prominent in La Niña years compared to the El Niño years. The whole central Indian belt from the desert regions of Rajasthan to the eastern coast shows a decline in rainfall, temperature increase, and hence, soil moisture declines in the La Niña years (Extended Figure 6). The dry western region shows a decline in rainfall and soil moisture even for the El Niño years with an increase in temperature (Extended Figure 5). The further drying patterns of the arid region due to interlinking could be alarming and hence, needs to be addressed in the planning for interlinking. Central Indian regions show an improvement in the rain due to interlinking in El Niño years, which is good for the dry years. Overall, we found that the perturbed water management from the proposed interlinking can lead to changes in the spatial distribution of the Indian Summer monsoon and a systematic reduction of precipitation in many regions, including the dry arid regions.

Further, to see whether the reduction in precipitation in IRR experiment is related to land-atmosphere feedback from extra irrigation provided in IRR experiment, we apply TE to find causal connections from the LH of extra irrigated regions to the precipitation over the drying regions. We considered the differences between IRR and CTL Simulations for each year separately for the same. The results are shown in Figure 4. We use TE here, as we want to capture both direct and indirect connections from latent heat flux to precipitation in model simulations. Figure 4(a) shows 3 chosen regions (southern peninsula – region A, western India – region B, and a part of ganga basin – region C) out of areas where irrigation was applied and Figure 4(b) shows a few selected regions where drying was witnessed (Region 1 – central-eastern India, Region 2 – Central India, Region 3 – western India, and Region 4 – western Himalayas). We tried to find the causal links from the latent heat flux (LH, IRR-CTL) of Regions A, B, and C to precipitation (P, IRR-CTL) from Regions 1, 2, 3, and 4. The link thickness as well as the link labels written at the beginning of the link show the number of times the link was found statistically significant out of 22 years of simulations. The node names are written as the variable symbol followed by the region they belong to for example, P1 means precipitation over region 1 and LH_A means latent heat flux in region A. The presence of causal links from LH to P from all regions indicates the influence of irrigation at regions A, B and C on precipitation over regions 1,2,3, and 4 via land-atmosphere feedback. The most robust links (present during most number of years) from all three irrigated regions are towards P1. Link from LH_C to P1 has highest consistency and was found during all the years. Our irrigation feedback results are consistent with the earlier studies that show Ganga basin is the global land-atmosphere feedback hotspot[43,49,55,56]. . Extended Figure 7 shows similar results for El Nino (Extended Figure 7,a) and La Nina (Extended Figure 7,b) years. Considering a link as consistent if it was found in more than 50% of years (at least 3 out of 5 years), connection from the LH_A, LH_B, and LH_C towards P1, P4, and P2 remain consistent for both El Nino and La Nina years. While links to P3 are consistent during El Nino years, they were found less consistent from LH_B and LH_C during La Nina. The presence of a links here is a measure of consistency of land-atmosphere feedback, and cannot be attributed to relative strength of drying during El Nino and La Nina years. Our results show that the land-atmosphere feedback from irrigated areas, especially from the Ganga basin and southern peninsula, remains consistent during El Nino and La Nina years.

**Implications of Interlinking Projects**

India has a rapidly growing problem of water stress due to global warming, population growth, pollution, and change in land use. As per the Central Water Commission, Government of India, the current per capita availability of water in India is around 1400 cubic meters which is slated to reduce to around 1200 cubic meters by 2050 and a large portion of country is already classified as water stressed[57,58]. A large fraction of India's water resources is used for agriculture and the irrigation is practiced across the country. The demand for water will further increase with rapid intensification of agriculture. As water demand is rising rapidly, within the next 20 years, India might need most of its runoff to meet its urban and agricultural needs[59]. As a solution to this problem, India has planned river-interlinking projects to transfer water from surplus to deficit basins to cater to the water demand of growing population. The goal is to keep maximum possible water on land for utilization which originally used to reach oceans from river basins. The assumption behind such planning is that 1) the hydrology of the river basins is independent; 2) the feedback from interlinking will not affect the rainfall patterns.

Here, we find that both the assumptions made for the interlinking are not valid. The perturbed hydrological processes of the receiving river basins send feedback to the Indian monsoon and potentially change the spatial patterns, specifically in the month of September. Such changing patterns of monsoon, in turn, affect the hydrology of the neighboring basins. Hence, in a true sense, the hydrological processes across river basins are not independent, a critical result which most large scale hydrological projects across the globe, including riverlinking projects in India do not consider while planning. The interdependence of river basins may have significant implications on water demand-availability tradeoff within a basin. Our WRF-CLM4 simulations are an attempt to quantify the possible changes in the Indian monsoon due to the proposed interlinking project. The results from these simulations show a systematic reduction of mean September precipitation of up to 12% in western arid region (states of Rajasthan and Gujarat), central (state of Madhya Pradesh), central-eastern (states of Odhisa and Chhattisgarh) and norther (states of Punjab, Haryana, and Uttarakhand) parts of India, which, based on our experimental set up, can be attributed to land-atmosphere feedback from interlinking. The reduction in September precipitation will dry up the rivers in the subsequent months amplifying water stress manifolds in various parts of the country, which is an unexpected and unintended result from interlinking. The majority of the population in these areas is dependent on agriculture. A reduction in monsoon rainfall would cause damages to socio-economy of these regions increasing climate vulnerability and risk. It is noteworthy that we have not considered the feedback on the monsoon rainfall in response to the reduced runoff to ocean due to interlinking.

Recent studies show land to ocean runoff can perturb the monsoon rain[60] and may intensify the feedback quantified by us. Hence, proper quantification of the feedback from proposed interlinking policy needs careful scientific investigation. Our study is the first attempt to quantify the impacts of any large scale hydrological project like river interlinking on Indian Summer Monsoon, which was not considered in planning these projects. Our results highlight the importance of regional land-atmosphere model-driven hypothesis testing and impacts assessment while planning for large-scale hydrological projects.

## Data and Methods

### Data

We use 40 years (1980-2019) of daily data from two sources (Extended Table 1): ERA-5 and MERRA-2 reanalysis products provided by the European Center for Medium-Range Weather Forecast (ECMWF, ERA-5)[20] and Modern-Era Retrospective analysis for Research and Applications Version 2 (MERRA-2)[21], respectively, over the major river basins of India (Figure 1(a)), Ganga (G), Godavari (Go), Mahanadi (M), Krishna (K), Narmada-Tapi (NT – two river basins taken together), Cauvery(C). Since Narmada and Tapi are small basins, we club them together and represent them as a single basin. Variables considered are Latent Heat Flux (LH), Sensible Heat flux (SH), Precipitation (P), Relative Humidity (R), Wind Speed (WS) (resultant of u-wind (U), v-wind (V)), Incoming Shortwave Radiation (SR), Soil Moisture up to root zone (SM) and Temperature (T). Variables R, U, and V are taken at 850 hpa pressure level and remaining variables are near surface. Variables were spatially averaged over each basin (Figure 1(a)) and thus generated 48 time series (8 variables on 6 basins) were converted to anomalies as deviations from their climatological mean values of 40 years. We use Oceanic Nino Index (ONI) values from National Oceanic and Atmospheric Administration (NOAA, https://origin.cpc.ncep.noaa.gov), USA, to classify El Nino and La Nina years as October to December ONI values of any year being above 1 and below -1 respectively (Extended Table 2).

**Transfer Entropy**

Information exchange takes place between two variables ($X_t$ and $Y_t$ with time series $x_1, x_2, \ldots, x_t$ and $y_1, y_2, \ldots, y_t$), when a change in one variable leads to a change in another. This information exchange gets reflected as an overlap of Shannon's Entropy[61] of observed time series of those variables. Binning any time series $X_t$ into m discrete parts, Shannon's Entropy can be computed as follows

$$H(X_t) = \sum_{i=1}^{m} p_i(x_t)(\log(p_i(x_t)))$$

Here $p_i(x_t)$ is probability of $x_t$ being in bin $i$. Transfer Entropy is an information-theoretic tool to delineate asymmetric connections in a non-linear dynamical system[16]. It is widely used in studying eco-hydrology and climate systems and is argued to be a well-suited measure of causality for such system[17,62–66]. It measures information from source variable to target variable while conditioning on the past of target variable. Since TE measures overlap of Shannon's Entropy, it can be estimated using $H(X_t)$ as follows

$$T(X_t > Y_t, \tau) = H(X_{t-\tau \Delta t}, Y_{t-\tau \Delta t}) + H(Y_t, Y_{t-\tau \Delta t}) - H(Y_{t-\tau \Delta t}) - H(X_{t-\tau \Delta t}, Y_t, Y_{t-\tau \Delta t})$$

Where, $T(X_t > Y_t, \tau)$ is Transfer Entropy from $X_t$ to $Y_t$ at lag $\tau$, $H(X_{t-\tau \Delta t}, Y_t, Y_{t-\tau \Delta t})$ and $H(X_t, Y_t)$ are the joint entropies between variables computed using joint probabilities instead of marginals. TE can be normalized by entropy of a distribution with m number of bins where all bins are equally likely $(H_{max} = \log(m))$. This form of TE estimation as previously been used in literature[17,62,67,68]. In this study, number of bins taken is 11 which has been argued to be appropriate for measuring TE given sufficient data length[17], time step $\Delta t$ is 1 (daily data) and lag $\tau$ varies from 1 to 10. After computing TE we test it for statistical significance at 95% confidence using method of shuffled surrogates[17,69–71].and compute normalized TE using $H_{max}$.

**PCMCI**

Peter and Clark's Momentary Conditional Independence (PCMCI) algorithm belongs to a class of causal discovery methods called '*Causal Network Learning Algorithms*' which first assume a fully connected causal graph and then iterate through each link testing for its removal by conditioning[14,15,72]. While TE faces trouble with high dimensional data, PCMCI handles the problem of high dimensionality[73] by dividing the process into two stages:

1. PC Stage reduces the dimensionality of conditioning set by filtering out less likely conditions. It iterating through all variables leading to a set of selected links called parents. Given $\overline{X}_t$, the set of all variables, for each variable $\overline{X}_t^j \in \overline{X}_t$, after initializing preliminary parents $\overline{\mathcal{P}}(X_t^j) = (\overline{X}_{t-1}, \overline{X}_{t-2}, \ldots, \overline{X}_{t-\tau_{max}})$ the following hypothesis is tested for all variables $\overline{X}_{t-\tau}^i$ from $\overline{\mathcal{P}}(X_t^j)$:

$$PC: X_{t-\tau}^i \perp\!\!\!\perp X_t^j \mid S$$

   for any set $S$ with cardinality $p$. Where $S$ contains a subset of $\overline{\mathcal{P}}(X_t^j) \setminus \{X_{t-\tau}^i\}$. We keep on increasing $p$ and test the null hypothesis which if we fail to reject, the link is removed from $\mathcal{P}$. The parents that remain in each iteration are sorted according to the value of the test statistic in decreasing order of strength and the next iterations start with the strongest parent first. A lenient alpha level of $\alpha = 0.2$ is taken for hypothesis testing in this stage so that true links are not lost.

2. MCI Stage finds causal connections for every pair $X_{t-\tau}^i \rightarrow X_t^j$ among above generate parents and time delays $\tau = \{1, 2, \ldots, \tau_{max}\}$ and tests the null hypothesis at $\alpha = 0.05$

$$MCI: X_{t-\tau}^i \perp\!\!\!\perp X_t^j \mid \overline{\mathcal{P}}(X_t^j) \setminus \{X_{t-\tau}^i\}, \overline{\mathcal{P}}(X_{t-\tau}^i) \,\forall\, X_{t-\tau}^i \in X_t^-$$

   Where $X_t^- = (X_{t-1}, X_{t-1}, \ldots, X_{t-\tau_{max}})$ and $\overline{\mathcal{P}}(X_{t-\tau}^i)$ is the conditioning set generated in the above step.

Both stages (PC and MCI) use conditional independence tests. In this study we use Conditional Mutual information using the k-nearest neighbor approach (CMI-knn)[74] as our choice of test statistic and maximum lag, $\tau_{max}$, taken for PCMCI is 10 days.

**WRF**

To simulate the surplus irrigation provided by river interlinking over the Indian region, we use a regional climate model - Weather Research and Forecasting model version 3 coupled with Community Land Model version 4 (WRF-CLM4)[18]. We use a modified the irrigation module in CLM4 that better represents the Indian practices of irrigation by incorporating ground withdrawal and flood irrigation practiced over paddy fields[9]. The domain used is from 59.5°E to 107°E and 3.7°S to 41.5°N (Extended Figure 3, f) with the model configured at 25 km grid spacing and 30 pressure levels in the vertical. We use initial and lateral boundary conditions from European Centre for Medium-Range Weather Forecast Interim Re-Analysis (ERA-Interim)[75] to perform two sets of experiments for 22 years (1991-2012) of the Indian monsoon season (ISM; 15 May to 31 October). Control experiment (CTL) prescribes current irrigation water application over India using estimates from the agricultural census and a gridded reconstructed data[52]. The Irrigation experiment (IRR) adds additional water as irrigation by maximizing the irrigated area fractions on the grids which are going to be benefitted from interlinking. Figure 1(b) shows (highlighted) the approximate grids to receive the additional irrigation from the river interlinking project[2,76,77] with the increase in the percentage irrgatead area from CTL to IRR run for each grid. Each grid cell projects an area of around 50000 hectares and overall, our IRR run contains an extra irrigated area of around 30 million hectares out of which around 20 million hectares belongs to regions fed by Himalayan rivers while around 10 million hectares belongs regions fed by peninsular rivers. The relative proportion of area irrigated from Himalayan and peninsular rivers is as per the interlinking plan by the government of India[3]. The amount of irrigation water added on these grids is estimated from the water availability and proposed culturable command area for the Kharif season from various projects[4] and is approximately equal to 600mm (around 4mm per day) for normal crops and 1450 mm (12 mm per day) for paddy.

A spin up time of 16 days (15 May to 31 May) is used in model runs to make sure the outputs are independent of initial conditions. For the choice of convective parameterization in WRF, we use Betts-Janjic-Miller (BMJ) scheme which is known to correlate well with spatial distribution of Indian Monsoon at daily scale[9]. To represent the sub-grid microphysics, planetary boundary layer, longwave and shortwave radiation, we use Lin Scheme, Yonsei University Scheme (YSU), Rapid Radiative Transfer Model (RRTM) and Dudhia scheme respectively. The parameterization is kept same for all simulations and the only difference between the two experiments (IRR and CTL) is the presence of additional irrigation in the IRR run to analyze changes in the simulated hydrometeorological variables during ISM.

**Acknowledgements**

We thank Sachin Budakoti and Roshan Jha for their technical discussions while configuring the regional model. The work is financially supported by the Department of Science and Technology Swarnajayanti Fellowship Scheme, through project no. DST/ SJF/ E&ASA-01/2018-19; SB/SJF/2019-20/11, and Strategic Programs, Large Initiatives and Coordinated Action Enabler (SPLICE) and Climate Change Program through project no. DST/CCP/CoE/140/2018. TC acknowledges the support from PM Research Fellowship, Government of India (RSPMRF0262), for his PhD scholarship.

**Data Availability**

All the datasets used in this study are publically available. ERA 5 reanalysis can be downloaded from https://cds.climate.copernicus.eu/cdsapp#!/dataset/reanalysis-era5-land. MERRA 2 reanalysis can be downloaded from https://gmao.gsfc.nasa.gov/reanalysis/MERRA-2/. Oceanic Nino Index (ONI) values are taken from https://origin.cpc.ncep.noaa.gov.

**Code Availability**

To perform PCMCI, a publically available python package 'tigramite' (https://github.com/jakobrunge/tigramite) was used. Code from regional climate model WRF-CLM4 modiefied to incorporate India specific irrigation is available at github https://github.com/IMMM-SFA/WRF_CLM4_Irrigation. The co-authors of the present manuscript prepared this India specific module.

**Author Contributions**

SG conceived the idea and designed the problem. TC and SG designed the hypothesis testing method. TC and AD performed all the analyses. MKR helped in the simulations. TC and SG primarily analyzed the results. KA and MKR reviewed the results and provided suggestions. SG and TC wrote the manuscript. All the authors reviewed the manuscript.

**Competing interests**

The authors declare no competing interests.

# Figures

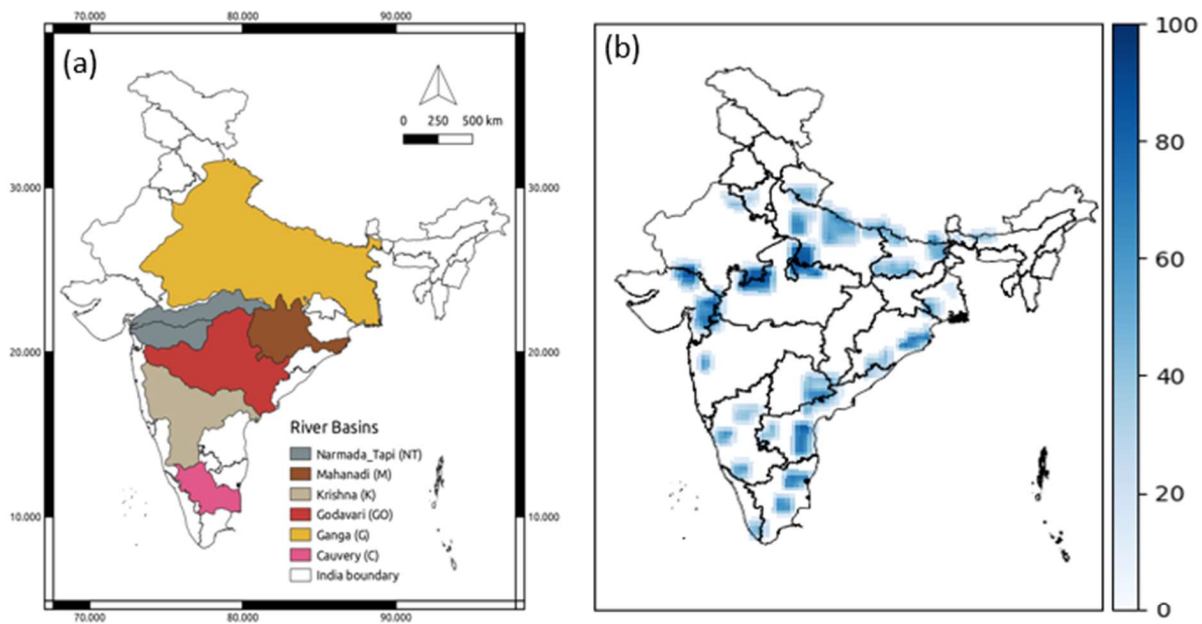

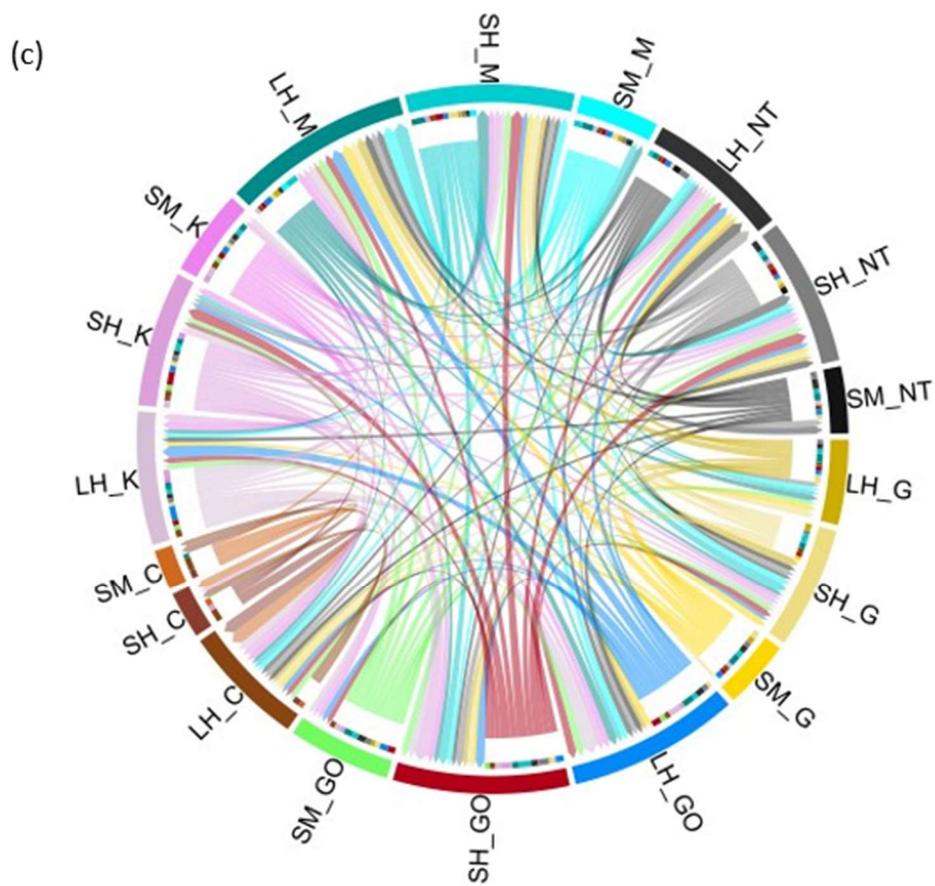

**Figure 1:** (a) River Basins in India considered in the study. (b) Irrigated grid cells under river interlinking schemes showing change in percentage irrigated area from CTL to IRR run (see methods) to increase irrigated area fraction to 80%. (c) Transfer entropy network between land variable across river basins. Various sectors are labelled as variable symbols (soil moisture-SM, latent heat flux-LH, sensible heat flux-SH) followed by the basin they belong to (Ganga (G, 808334 km$^2$), Godavari (Go, 302063 km$^2$), Mahanadi (M, 139659 km$^2$), Krishna (K, 254743 km$^2$), Narmada-Tapi (NT, 98,796 km$^2$, 65,145 km$^2$ respectively – two river basins taken together), and Cauvery(C, 85624 km$^2$)). Links are only shown if found statistically significant at 95% confidence and are colored same as the node they originate from. For example, link from LH_G to LH_C shows that there is a connection between latent heat fluxes from Ganga and Cauvery basin. Ratio of incoming to outgoing links in cauvery basin is very high compared to ganga basin.

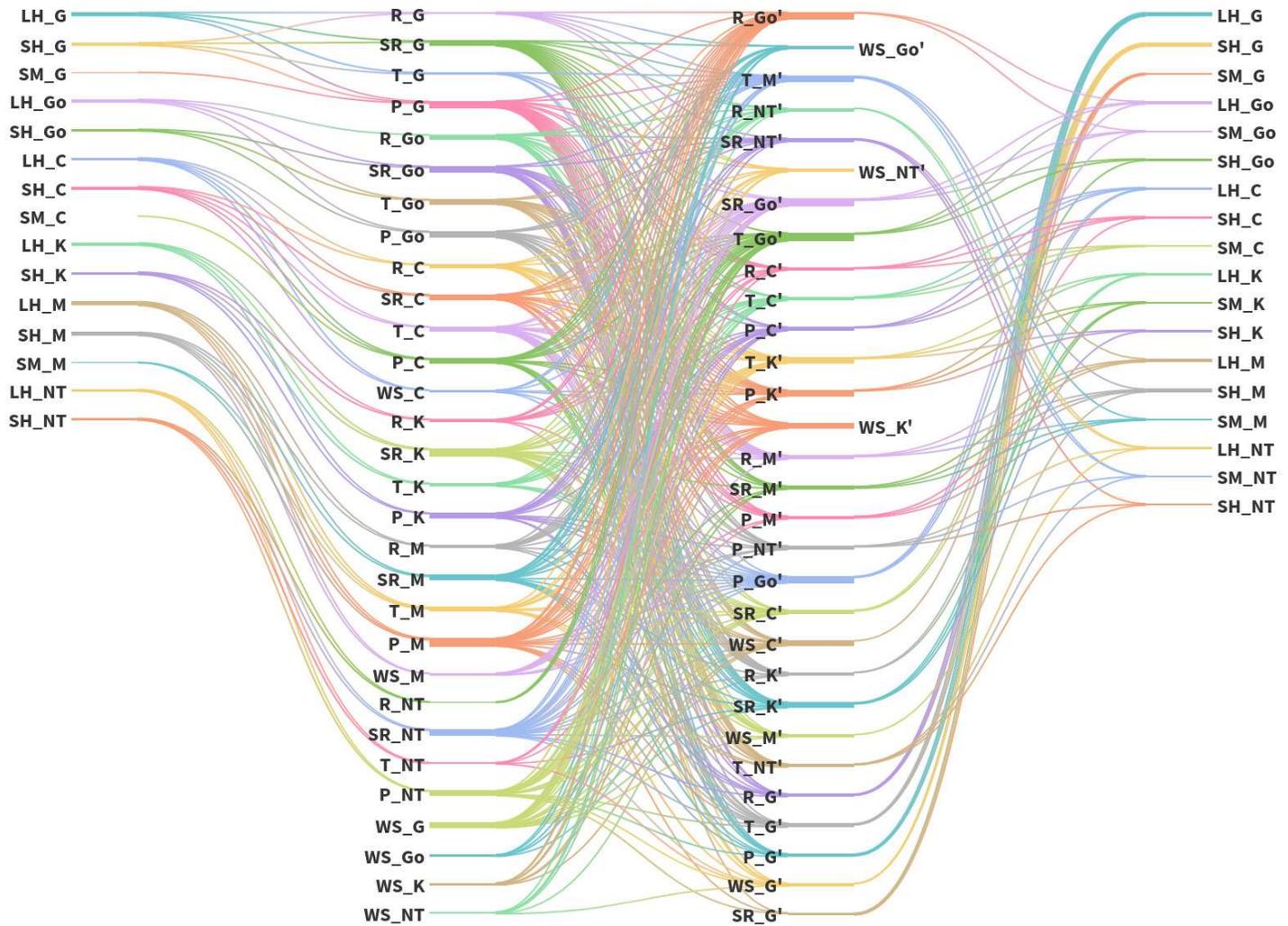

**Figure 2:** Connections using PCMCI from Land to atmosphere within basin (first column), between atmospheric variables across all basins (second column), and from atmosphere to land within the basin (third column). A link is shown only if it appeared more than 25% of the time (10 years out of 40 years (1981-2020)). Names are variable symbols followed by the basin they belong to for example, LH_G means latent heat flux from Ganga basin. The first and second column of variables are land variables (soil moisture SM, latent heat flux LH, and sensible heat flux SH) and atmosphere variables (precipitation P, temperature T, relative humidity R, wind speed WS, and incoming short wave radiation SR) respectively, links between which represent land-to-atmosphere connections within each basin. Links between next two columns represent atmosphere to atmosphere connections, for example, there is a link from temperature in Ganga basin (T_G) to that of Mahanadi basin (T_M'). Links in the last

column represent downward connections from atmospheric variables to land variables within basin.

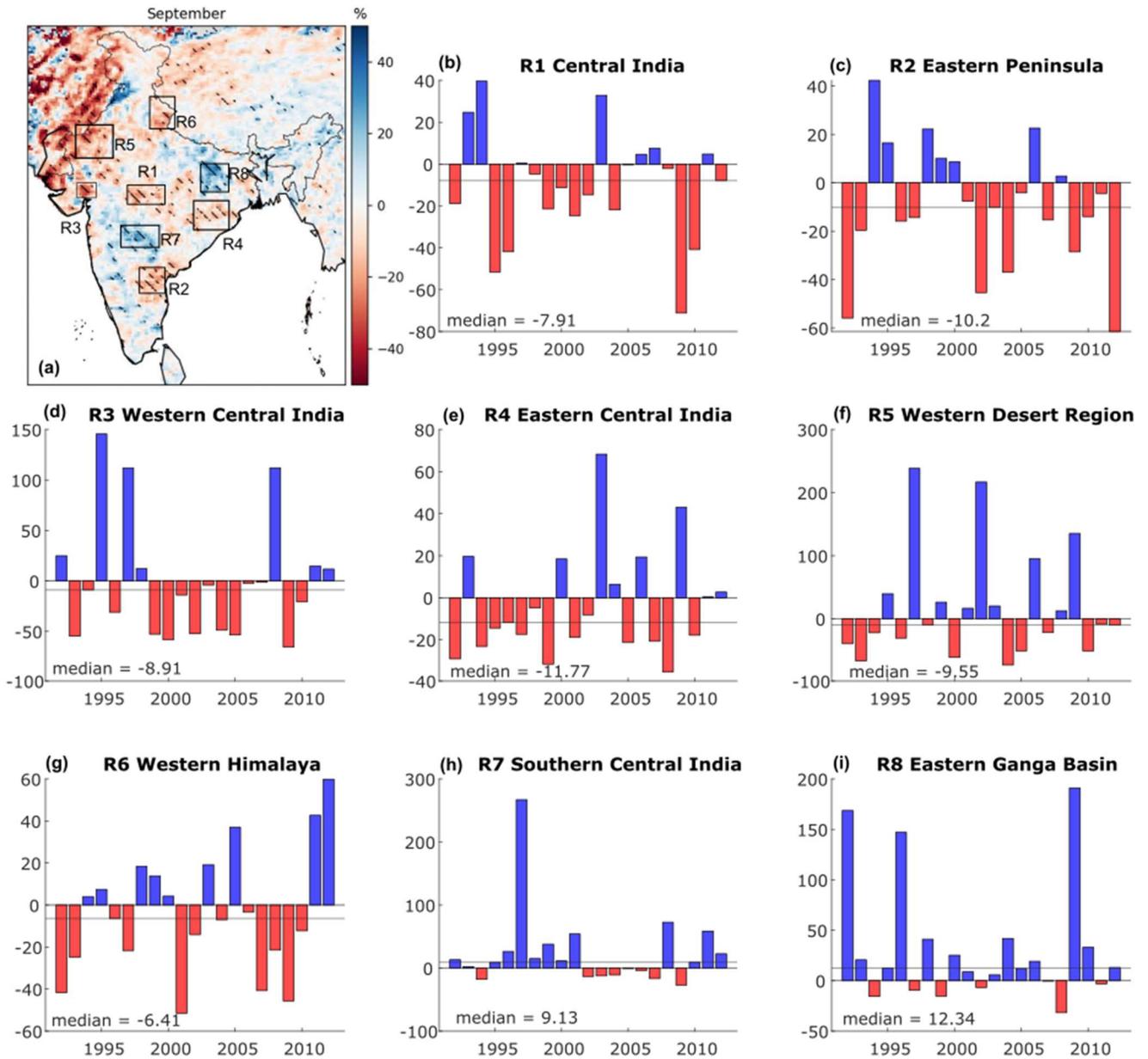

**Figure 3:** (a) Percentage change in mean daily precipitation between WRF runs IRR and CTL (IRR-CTL) for the month of September. Hatch lines mark regions where the difference was found statistically significant at 90% confidence tested on 660 data points. (b-i) Percentage change (red: decrease, blue: increase) for all years in mean September rainfall for regions marked in (a). Median change for all years is represented by horizontal line and is also written as text in each plot. There is a significant reduction in September precipitation of upto 13% in central (state of Madhya Pradesh), eastern (states of Odhisha and Chhattisgarh), northern (state of Uttarakhand in western Himalaya), and western arid region (states of Rajasthan and Gujarat) of India.

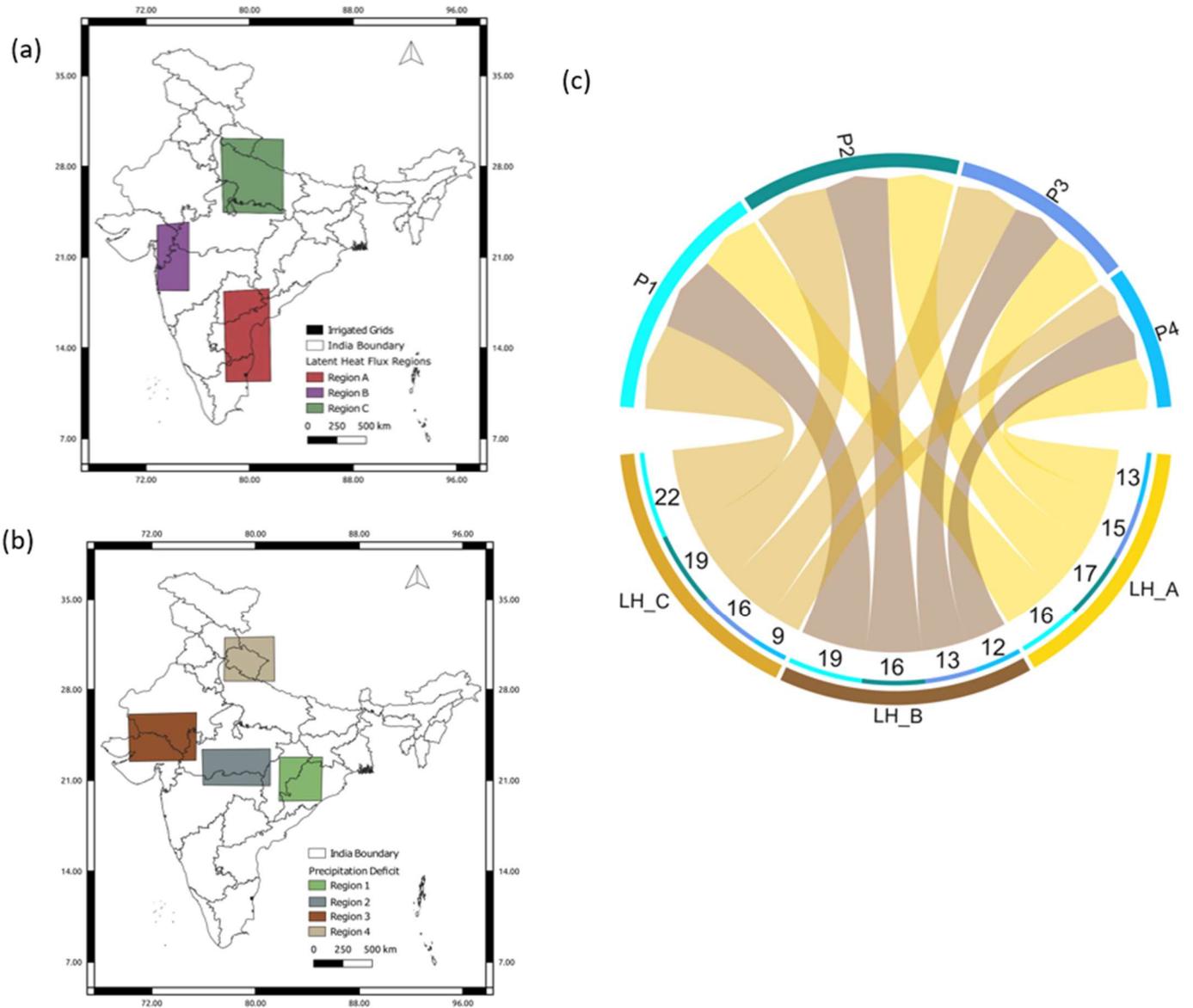

**Figure 4:** (a) Selected regions where irrigation was applied. (b) Selected regions of September precipitation where drying is observed due to interlinking. (c) Connections form change in latent heat flux (IRR-CTL) in the irrigated regions to change in precipitation (IRR-CTL) in the highlighted regions using Transfer Entropy. Links are labelled as the number of years, when they were found out to be statistically significant out of 22 years of WRF simulation (1991-2012). This shows that change in precipitation from CTL to IRR experiment is causally related

to the corresponding change in latent heat flux of other regions indicating consistent land-atmosphere feedback.

# Extended Figures

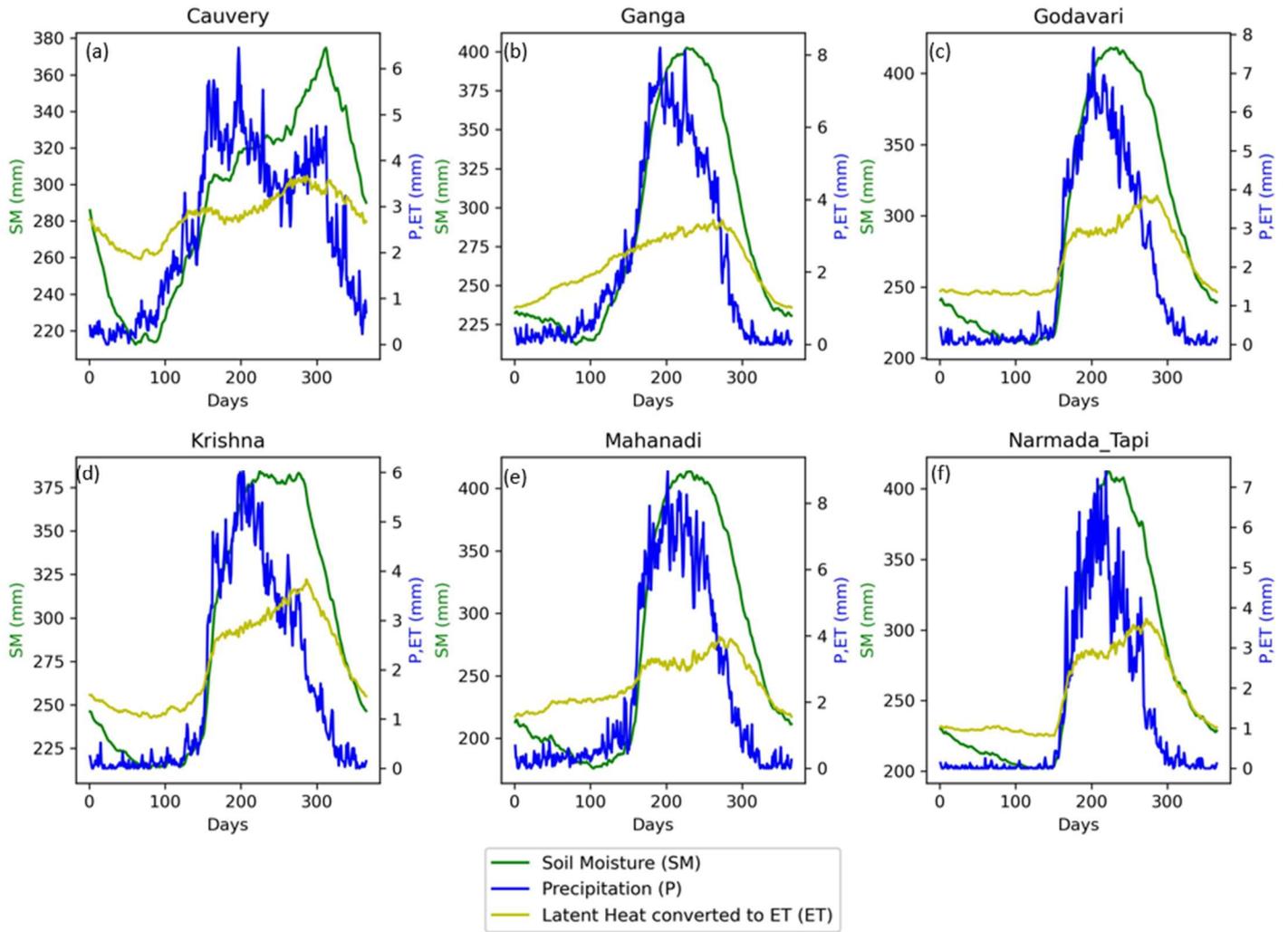

**Extended Figure 1:** Climatology of soil moisture (SM), precipitation (P), and evapotranspiration (ET) in different basins, as obtained from the ERA-5 reanalysis.

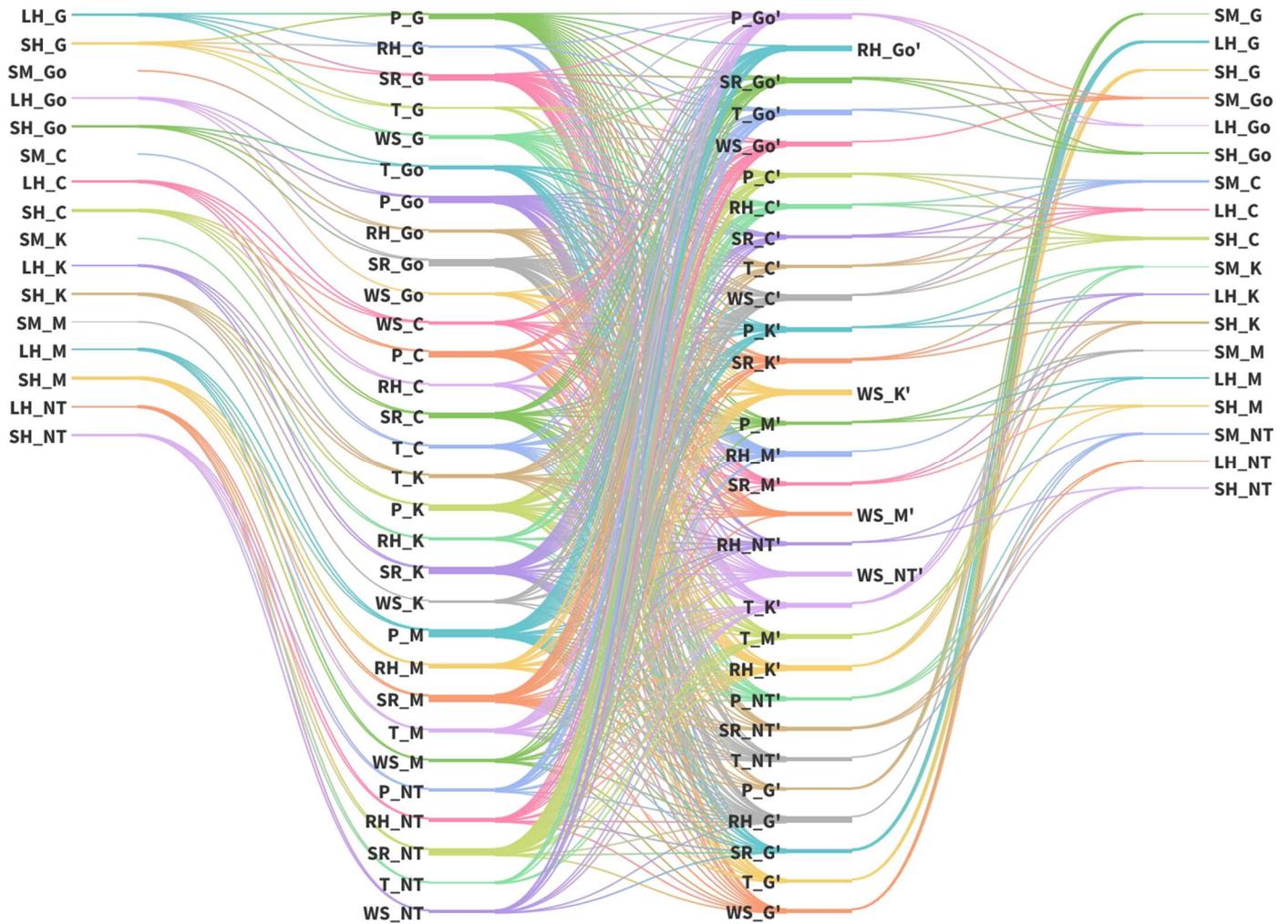

**Extended Figure 2:** Causal connections of land atmosphere interactions from MERRA-2 Reanalysis using PCMCI (similar to Figure 2). A link is shown only if it appeared more than 25% of the time (10 years out of 40 years (1981-2020)).

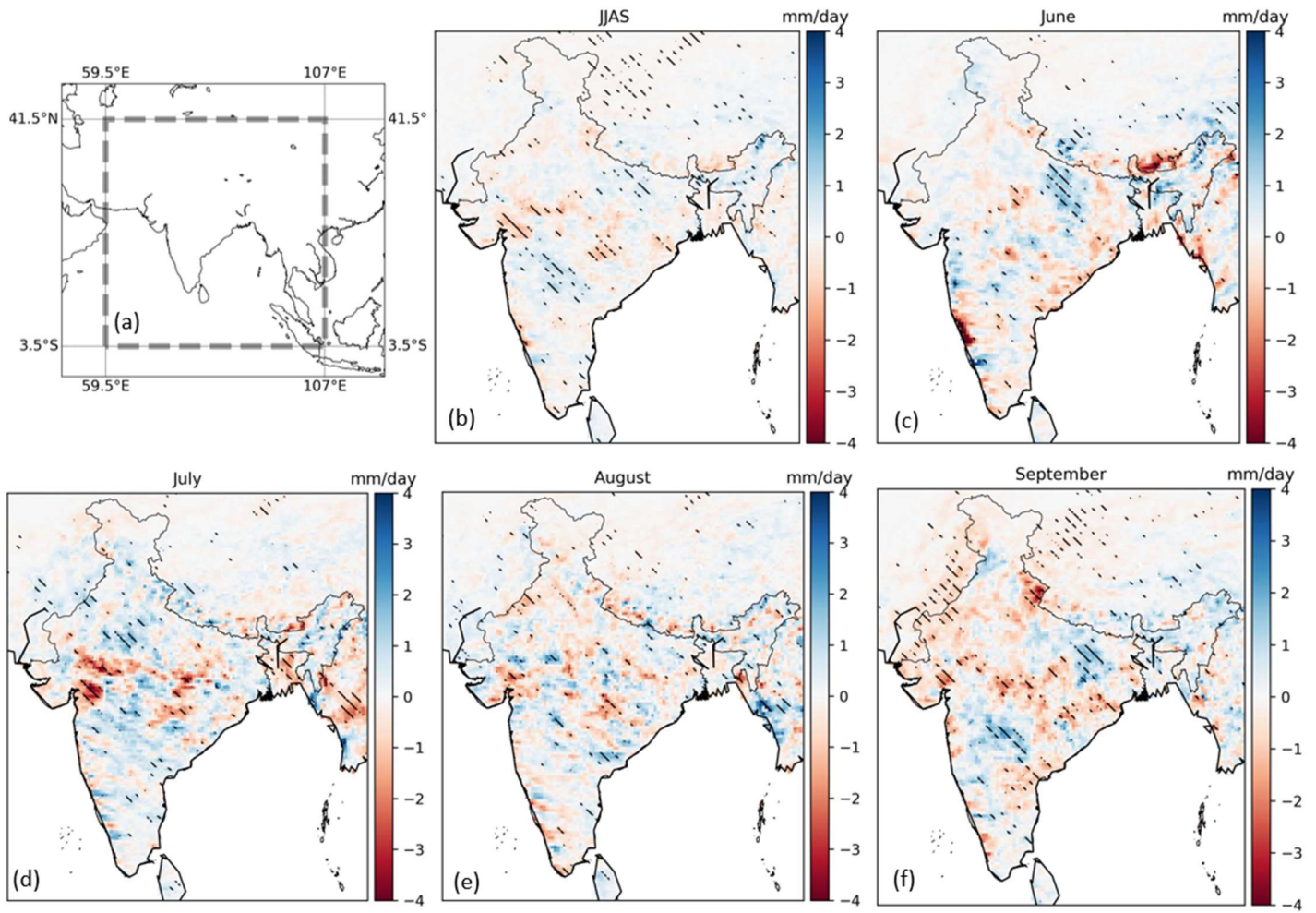

**Extended Figure 3:** (a) Domain used for WRF simulations. (b-e) The difference in mean Precipitation (mm/day) between IRR and CTRL runs from WRF for the complete monsoon season (b, JJAS; June to September) and individual months June (c), July (d), August (e), September (f). Hatched regions indicate statistically significant differences at 90% confidence tested on 660 data points.

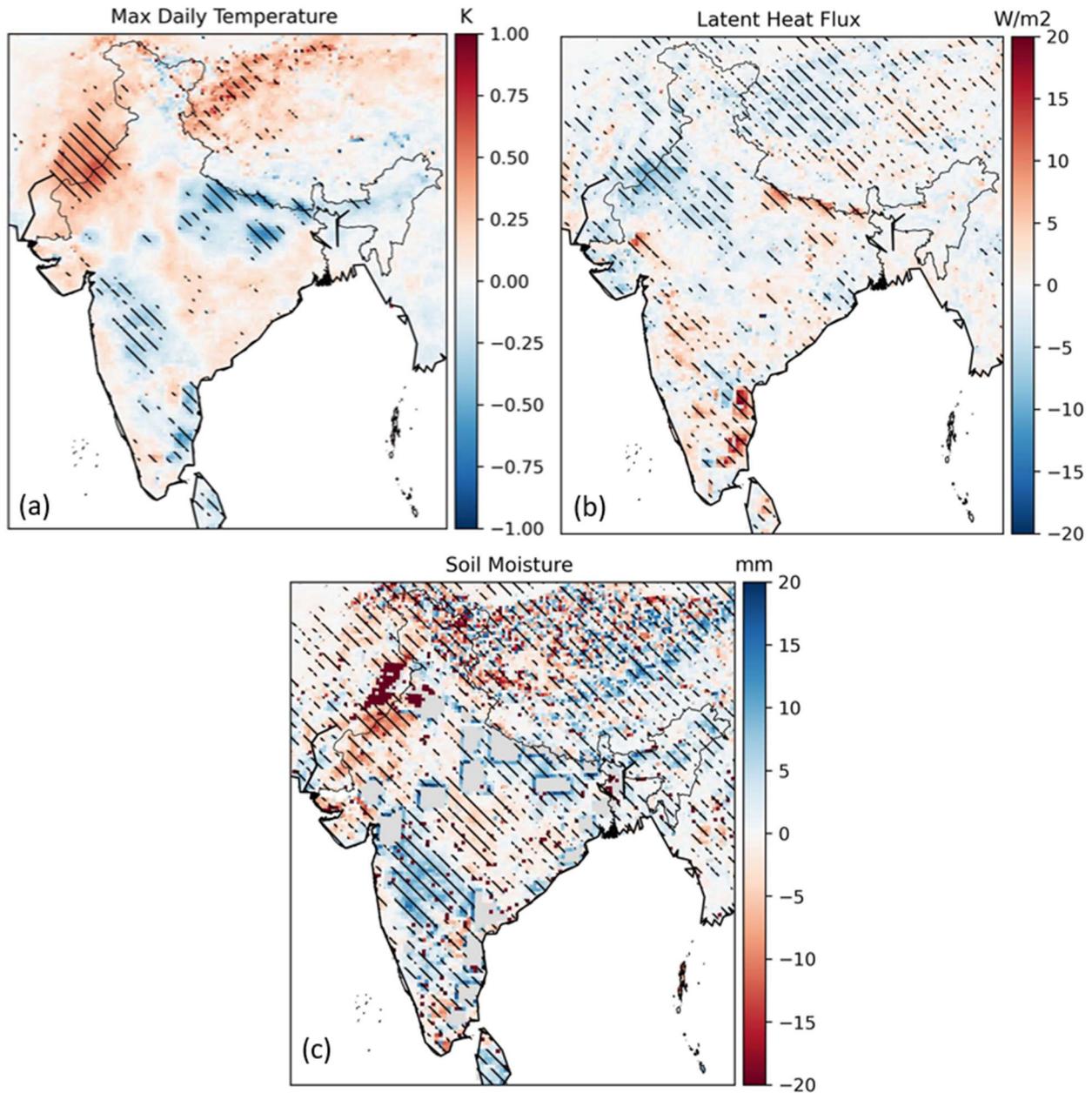

**Extended Figure 4:** (a-c) Differences between mean values from IRR and CTL runs from WRF during the month of September for variables daily maximum temperature (a; K), mean daily latent heat flux (b; W/m2) and mean monthly soil moisture (c; mm). Hatched regions represent statistically significant grids at 90% confidence tested on 660 data points.

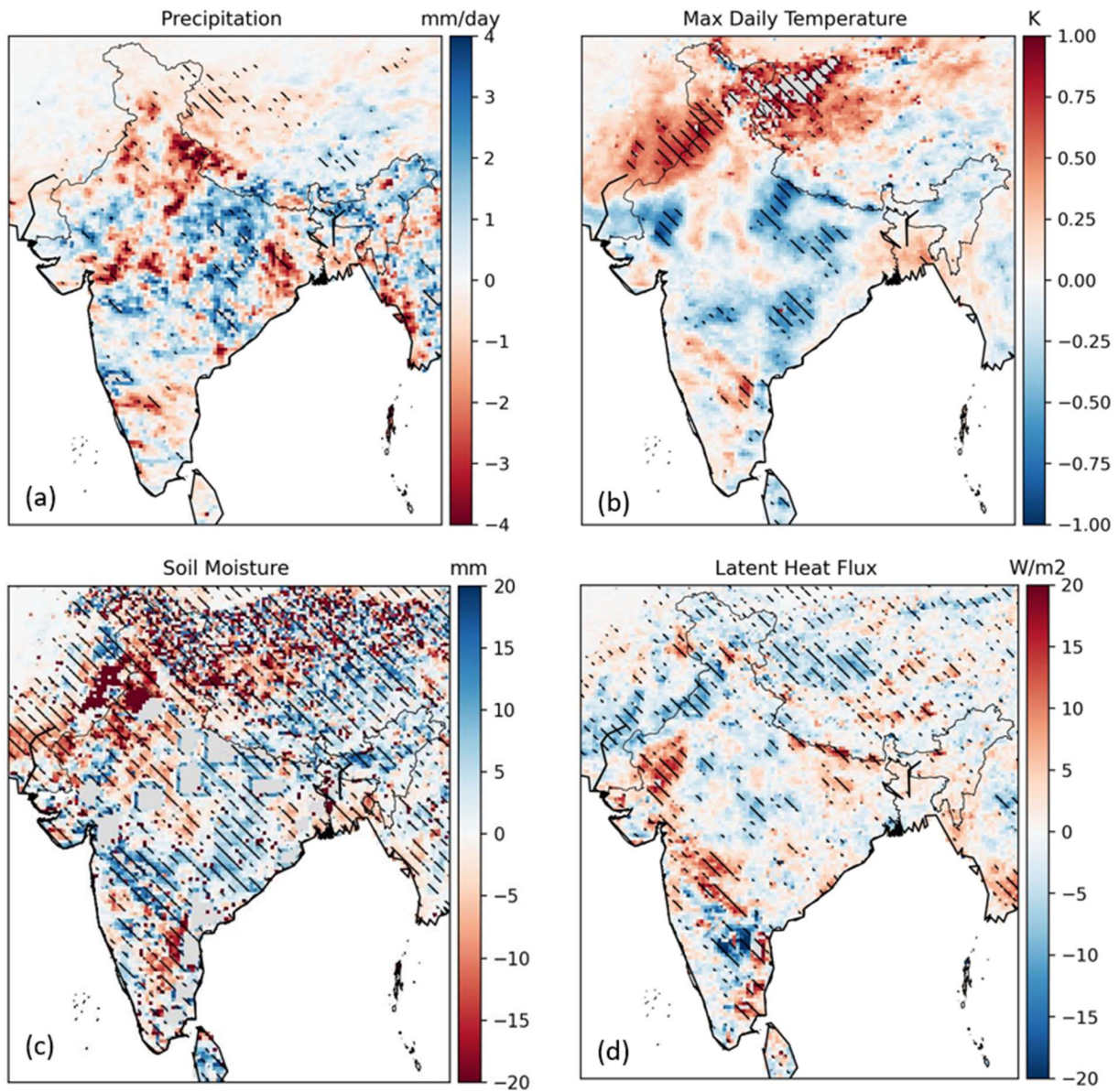

**Extended Figure 5:** Difference between mean values from IRR and CTL runs from WRF during September months of El Nino Years for daily precipitation (a; mm/day) , daily maximum temperature (b; K), mean monthly soil moisture (c; mm) and mean daily latent heat flux (d; W/m2). Hatched regions represent statistically significant grids at 90% confidence tested on 150 data points.

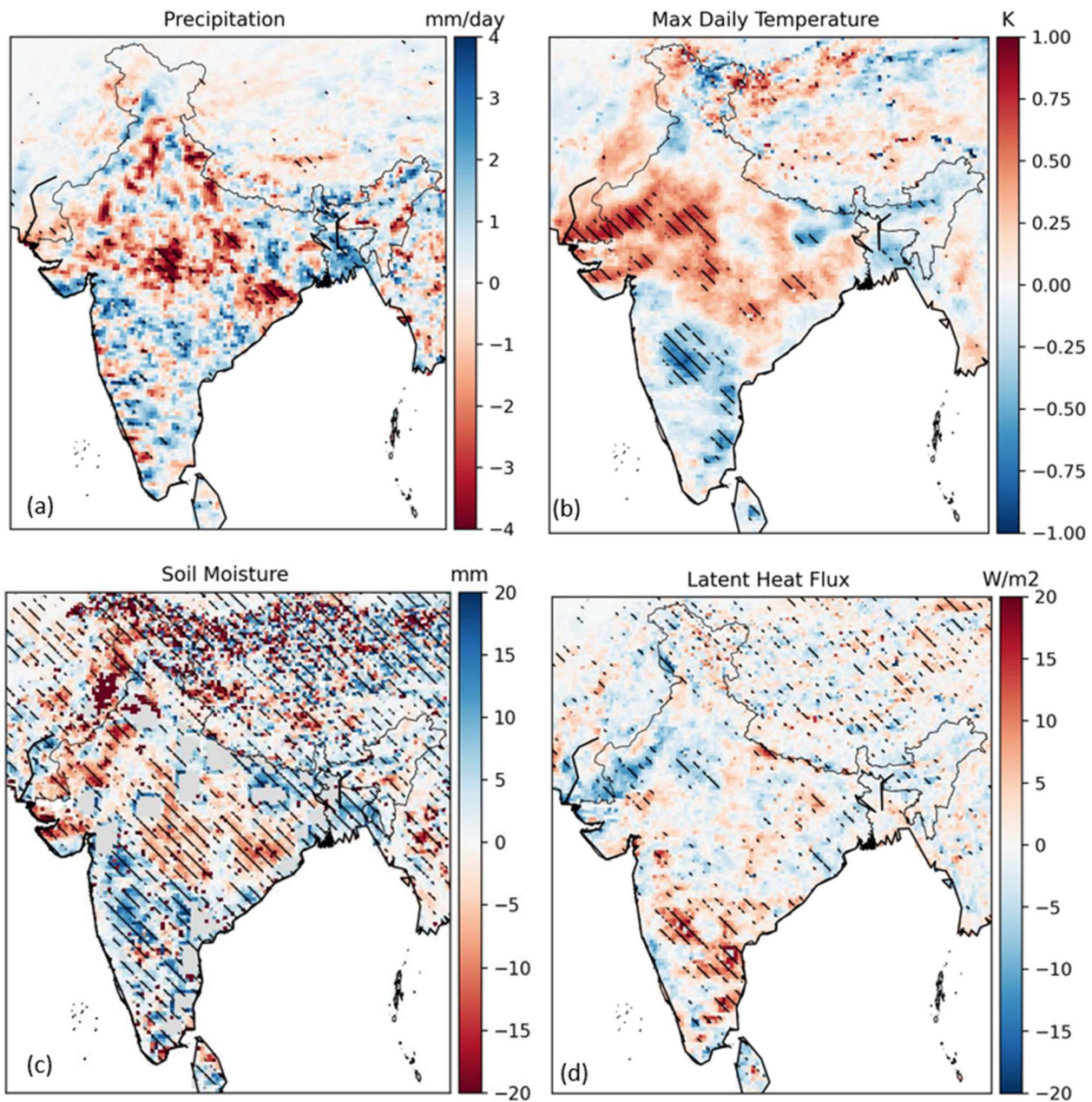

**Extended Figure 6:** Difference between mean values from IRR and CTL runs from WRF during September months of La Nina Years for daily precipitation (a, mm/day), daily maximum temperature (b, K), mean monthly soil moisture (c, mm), and mean daily latent heat flux (d, W/m2). Hatched regions represent statistically significant grids at 90% confidence tested on 150 data points.

**Extended Figure 7:** Transfer Entropy form changes (IRR-CTL) in latent heat flux of irrigated regions to precipitation change (IRR-CTL) in drying regions for El Nino years (a) and La Nina years (b). Links are labelled as the number of years, when they were found out to be statistically significant out of 5 El Niño and 5 La Niña years (Extended Table 2). This shows the consistency of land-atmosphere feedback during both El Niño and La Niña years.

# Extended Tables

**Extended Table 1:** Variables considered over each basin.

| Data | Symbol | Spatial Resolution | Source |
| --- | --- | --- | --- |
| Precipitation | P | 0.25° X 0.25°/0.5° X 0.667° | ERA-5/MERRA-2 |
| Soil Moisture | SM | 0.25° X 0.25°/0.5° X 0.667° | ERA-5/MERRA-2 |
| Latent heat flux from land | LH | 0.25° X 0.25°/0.5° X 0.667° | ERA-5/MERRA-2 |
| Sensible heat flux from land | SH | 0.25° X 0.25°/0.5° X 0.667° | ERA-5/MERRA-2 |
| Shortwave flux incoming on land | SW | 0.25° X 0.25°/0.5° X 0.667° | ERA-5/MERRA-2 |
| Relative Humidity at 850 hpa | Q | 0.25° X 0.25°/0.5° X 0.667° | ERA-5/MERRA-2 |
| U-Wind at 850 hpa | U | 0.25° X 0.25°/0.5° X 0.667° | ERA-5/MERRA-2 |
| V-Wind at 850 hpa | V | 0.25° X 0.25°/0.5° X 0.667° | ERA-5/MERRA-2 |
| 2 meter air Temperature | T | 0.25° X 0.25°/0.5° X 0.667° | ERA-5/MERRA-2 |

**Extended Table 2:** Years when Oceanic Niño index (ONI) exceeded 1 (El Niño) and was less than 1 (La Niña).

| El Niño | La Niña |
| --- | --- |
| 1991 | 1995 |
| 1994 | 1998 |
| 1997 | 1999 |
| 2002 | 2007 |
| 2009 | 2010 |